\documentclass[photonics,article,accept,pdftex,moreauthors]{Definitions/mdpi} 

\firstpage{1} 
\pubvolume{10}
\issuenum{_7}
\articlenumber{774}
\pubyear{2023}
\copyrightyear{2023}
\datereceived{25 May 2023} 
\daterevised{25 June 2023} % Comment out if no revised date
\dateaccepted{1 July 2023} 
\datepublished{ } 
\Title{Astronomical Camera Based on a~CCD261-84~Detector with Increased Sensitivity in~the~Near-Infrared}

\Author{Irina Afanasieva $^{1,}$*\orcidA{}, Valery Murzin $^{1}$, Valery Ardilanov $^{1}$\orcidC{}, Nikolai Ivaschenko $^{1}$, Maksim Pritychenko $^{1}$,  \linebreak Alexei Moiseev $^{1, 2}$\orcidD{}, Elena Shablovinskaya $^{1}$\orcidE{} and Eugene Malygin $^{1}$\orcidF{}}

\AuthorNames{Irina Afanasieva, Valery Murzin, Valery Ardilanov, Nikolai Ivaschenko, Maksim Pritychenko, Alexei Moiseev, Elena Shablovinskaya and Eugene Malygin }

\address{%
$^{1}$ \quad Special Astrophysical Observatory, Russian Academy of Sciences, Nizhniy Arkhyz 369167,  Russia; vamur@sao.ru (V.M.); valery@sao.ru (V.A.); ivanick@sao.ru (N.I.); pma@sao.ru (M.P.); moisav@sao.ru~(A.M.); shabli@sao.ru (E.S.); male@sao.ru (E.M.)%\\

$^{2}$ \quad Sternberg Astronomical Institute, Moscow M.V. Lomonosov State University, Universitetskij Pr. 13, \linebreak Moscow 119992, Russia}

\corres{Correspondence: riv@sao.ru}

\abstract{Herein, we describe the design, implementation and operation principles of an astronomical camera system, based on a large-format CCD261-84 detector with an extremely thick $200\,\upmu$m substrate. The DINACON-V controller was used with the CCD to achieve high performance and low noise. The CCD system photometric characteristics are presented. A spatial autocorrelation analysis of flat-field images was performed to reveal the dependence of substrate voltage on the lateral charge spreading. The investigation of the dispersion index for the optimal choice of exposure time is discussed. Studies of the patterns of fringes were carried out in comparison with previous detectors. The amplitude of fringes with CCD261-84 was significantly lower, compared to previous-generation detectors. The results of using a new camera for imaging and spectral observations at the Russian 6 m telescope  with the SCORPIO-2 multimode focal reducer are considered. The developed CCD camera system makes it possible to significantly increase the sensitivity in the 800--1000 spectral~range.}

\keyword{instrumentation; fully depleted scientific detectors; instrument optimisation; statistics; autocorrelation; imaging spectroscopy}  % CCD detector system; measurement;

\begin{document}

%%%%%%%%%%%%%%%%%%%%%%%%%%%%%%%%%%%%%%%%%%

% The order of the section titles is different for some journals. Please refer to the "Instructions for Authors” on the journal homepage.

\section{Introduction}

The near-infrared (NIR) band (from 750~nm to 2500~nm) is of considerable interest to astronomers \cite{Rieke2009}. To achieve maximum sensitivity in this range, astronomical camera systems mainly use matrix multiplexed hybrid IR (infrared) detectors. Such detectors are much more sensitive than silicon devices, but they are vastly more expensive and not widely available \cite{Vincent2016}.

For wavelengths up to 1100~nm, silicon CCD detectors have recently become an alternative to hybrid devices. Conventional CCD detectors have very high sensitivity in the visible range, but, at wavelengths greater than 700--800~nm, they show a significant decrease in sensitivity \cite{McLean2008}. This is due to the small thickness of the silicon device substrate, which leads to a decrease in the absorption of long-wave photons in silicon.

The progress in the development of silicon CCD detectors in the field of increasing sensitivity in the red and near-infrared bands can be observed by the example of improving the line of scientific detectors manufactured by Teledyne E2V~\cite{e2v2023}. Conventional backside illuminated (BSI) CCD detectors manufactured using epitaxial silicon technology with a substrate thickness of no more than $20\,\upmu$m have a sharp decrease in sensitivity starting from 700~nm~\cite{Hayes1997}. The use of a thicker substrate up to $40\,\upmu$m with deep depletion technology improves red sensitivity. The next class of devices with additionally improved IR sensitivity are CCD detectors with medium substrate thickness. These devices, based on bulk silicon with substrates up to $70\,\upmu$m, demonstrate high red sensitivity and good photometric characteristics~\cite{Downing2009}. The last level of development is devices with a thick substrate (more than $100\,\upmu$m) of high-resistance full depletion silicon, called High-Rho CCD detectors~\cite{Jorden2006, Robbins2011}. They have the highest sensitivity in the near-infrared range for their class~\cite{Jorden2010}.

%``\textit{High-Rho}''

In 2018, the Special Astrophysical Observatory of the Russian Academy of Sciences (SAO RAS) acquired the High-Rho CCD261-84, a full-frame scientific BSI CCD detector with a frame format of $2048\times4104$ and $200\,\upmu$m substrate thickness, manufactured by Teledyne E2V~\cite{Jorden2014}. The detector has very high sensitivity in the red and near-infrared bands. In this paper, we present the implementation of an acquisition system with this photodetector (in this paper, referred to as CCD System) and demonstrate the results of studying its photometric characteristics and the features associated with its operation on the 6 m Big Telescope Alt-azimuthal (BTA) of the SAO RAS as a part of the {SCORPIO-2} multimode focal reducer~\cite{SCORPIO2}.

%%%%%%%%%%%%%%%%%%%%%%%%%%%%%%%%%%%%%%%%%%
\section{CCD System Construction}

Over the past 20 years, SAO RAS has created 5 generations of CCD controllers able to manage all kinds of CCD detectors, including various mosaic configurations~\cite{Markelov2000, Murzin2016}. On the one hand, the creation of new generations of CCD controllers is caused by the appearance of new photosensitive devices and tasks to be solved, and by the emergence of new electronic components with qualitatively increased characteristics on the other hand.

DINACON-V is the latest generation of CCD controllers built on the basis of modern electronic components ~\cite{Murzin2016}. The controller architecture allows the handling of both single and mosaic photodetectors with a total number of video channels up to 256 and a maximum data throughput between the controller and the host computer up to 10~Gbit/s. One of the main features of the controller is the ability to implement camera systems using High-Rho CCD detectors, which require high control voltage to create a sufficient electric field on~a~thick substrate~\cite{Ardilanov2020}.

\begin{figure}[H]
\includegraphics[width=0.7\textwidth]{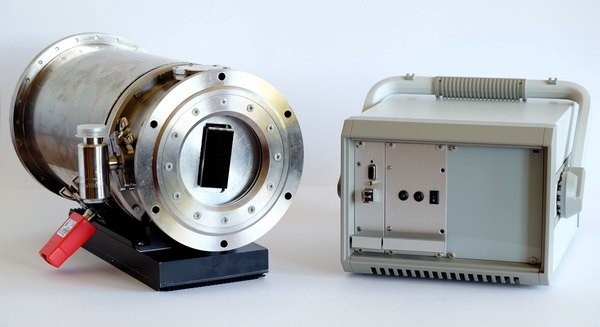}
\caption{General view of the CCD261-84 camera system: the camera (\textbf{left}) and the PSU~(\textbf{right}). \label{fig1}}
\end{figure}

The camera system with the CCD261-84 detector (Figure~\ref{fig1}) consists of two parts: a~nitrogen-cooled CCD camera and a power supply unit (PSU).

%\unskip

Structurally, the camera and the power supply unit can be separated from each other by up to~1.5~m, which is limited by the length of the connecting cables. The distance between the CCD camera and the host computer can be up to 200~m and is determined by the required length of the fiber optic communication line. The system is designed to operate in an ambient temperature range from~$-40$ to~$+40$~\textdegree~C.

%%%%%%%%%%%%%%%%%%%%%%%%%%%%%%%%%%%%%%%%%%
\section{CCD Camera Design}

\begin{figure}[H]
\begin{adjustwidth}{-\extralength}{0cm}
\centering
\includegraphics[width=1.2\textwidth]{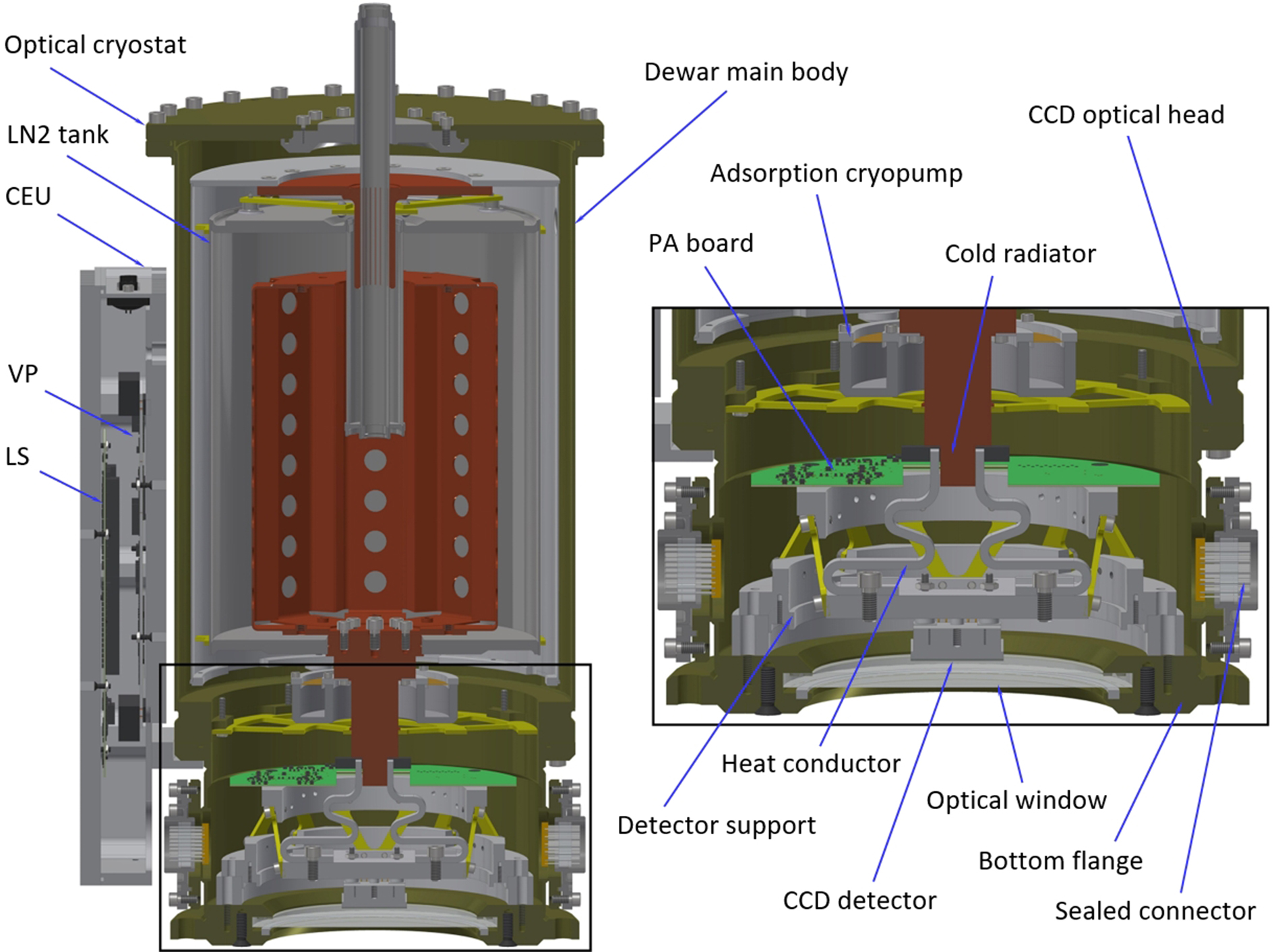}
\end{adjustwidth}
\caption{A 3D cut view of the CCD camera: full view of the camera (\textbf{left}) and zoomed view of the optical head (\textbf{right}). \label{fig2}}
\end{figure}

The left part of Figure~\ref{fig2} shows the design of the CCD camera, including an optical cryostat and a~camera electronics unit (CEU).

The optical cryostat consists of the liquid nitrogen (LN2) cryostat itself and the optical head (Figure~\ref{fig2}, right). The nitrogen cryostat includes a nonspillable type of nitrogen tank, an adsorption cryopump, and a cold radiator for connecting the heat load. The filler neck for filling liquid nitrogen is located at the top of the cryostat. The optical head includes a detector support with a CCD detector, a preamplifier (PA) board connected to the CCD detector, and a flexible heat conductor connected to a cold heatsink. 

The CEU includes a hermetic case, which also provides the removal of heat generated by the electronics placed inside the case, a video processor--generator board (VP), and a level shaper board (LS). The electrical connection between the camera electronics unit and the PA board is carried out through a sealed connector on the optical head.

The CCD camera has an input optical window and can be mounted at a telescope or spectrograph focus in an arbitrary spatial position using mounting holes in the bottom~flange.

%The retention time of LN2 in the operating mode is 16 h, which does not require refueling during the night.

%%%%%%%%%%%%%%%%%%%%%%%%%%%%%%%%%%%%%%%%%%
\section{CCD Controller Electronics}

The controller includes the PA board, the VP, the LS, and the communication module (CM) with the power source. The PA board is located in the CCD optical head, and the VP and LS are in the camera electronics unit, which is mounted directly on the optical cryostat. The CM with a power source is structurally placed in a separate block (Figure~\ref{fig1}, right) and connected to the CEU with two connecting cables. 

Placing a preliminary amplifier in the camera is necessary to amplify the video signal as close as possible to the outputs of the CCD detector. This approach enables one to achieve the maximum signal-to-noise ratio as well as to perform signal conversion to a differential form, which is necessary to minimize interference during video signal transmission and for subsequent analog-to-digital conversion (ADC). The PA board also includes electrostatic protection circuits and protection against increased operating voltages on the CCD detector. Inside the camera optical head, there is also a temperature sensor and photodetector base heater. The temperature sensor and heater are the elements of the photodetector temperature stabilization circuitry.

\begin{figure}[H]
\begin{adjustwidth}{-\extralength}{0cm}
\centering
\includegraphics[width=1.1\textwidth]{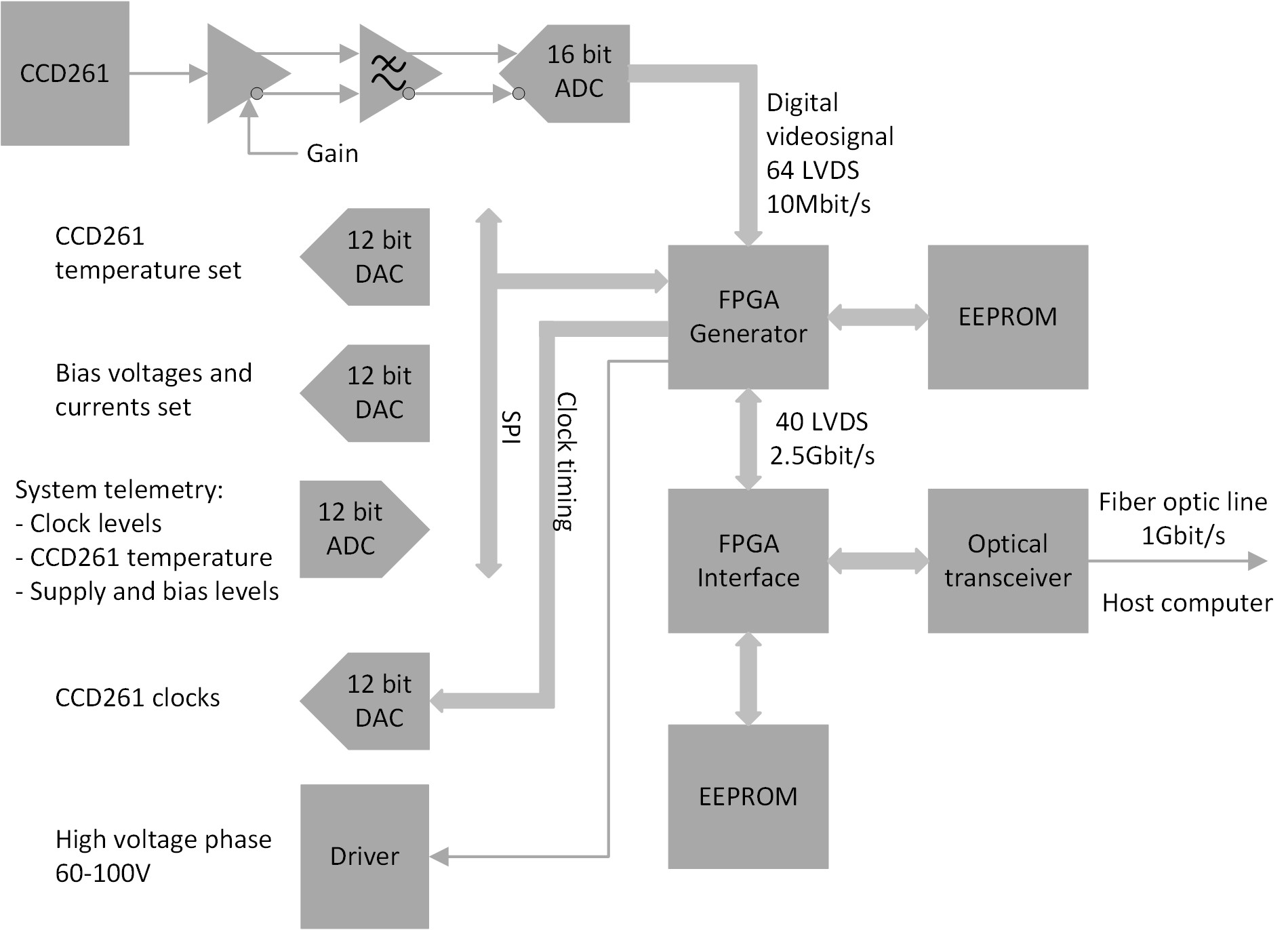}
\end{adjustwidth}
\caption{Block diagram of the DINACON-V CCD controller. \label{fig3}}
\end{figure}

The operation of the DINACON-V CCD controller can be explained using the block diagram shown in Figure~\ref{fig3}.

In the CEU, digital circuitry is implemented in a Field Programmable Gate Array (FPGA), and the analog part of the circuits is controlled by DAC (digital-to-analog converter) and ADC. The CEU is housed in a hermetically sealed enclosure and performs the following tasks:
\begin{itemize}
\item  Formation of control voltages and signals of the CCD detector and their telemetry;
\item  Stabilization of the operating temperature of the CCD detector;
\item  Processing of analog video signal with subsequent ADC and noise-reducing digital~filtering;
\item  Organization of an interface with the communication module for digital data exchange.
\end{itemize}

The generated timebase logic signals are then further converted to the required levels to control the CCD detector. In addition to standard voltage levels, a high-voltage level (up to $-$100\,V) is generated and applied to the back side of the CCD detector substrate. This voltage can be adjusted to minimize the effect of charge spreading.

CCD261-84 has two video outputs, and, accordingly, two video channels are implemented. The video signal processing channel uses components with low leakage currents, low noise and high speed. Due to the low leakage currents, signal clamping occurs once per line, which significantly improves performance compared to the need to clamp the signal in each pixel. 

The processing channel also has stages with gain switching and low-pass filtering. For further video signal conversion into digital form, a high-precision 10 MHz 16-bit ADC is used. The ADC--FPGA interface is organized by serial differential lines, which minimizes interference from digital circuits.

At the stage of digital processing and filtering of the video signal performed in the FPGA, three video signal readout rates---fast, normal or slow with 4, 16 or 64 signal samples, respectively, in one pixel---are implemented. These are processed by an antinoise optimal digital filter with a finite impulse response. The filter coefficients are determined in advance based on the noise spectrum measurement of the CCD output stage. The CCD controller provides not only very low read noise but also high stability of the video channel transfer characteristic under conditions of a large ambient temperature variation around the telescope, which is essential for photometry tasks.

The communication module synchronizes processes in the CCD controller and performs the following tasks: 
\begin{itemize}
\item  Receiving and executing commands from the host computer;
\item  Receiving video data coming from camera electronics and sending video and telemetry data to the control computer; 
\item  Exposure synchronization with external precise time service;
\item  External shutter control.
\end{itemize}

Data packet exchange between the CCD controller and the host computer occurs via a fiber optic interface using the Ethernet 1~Gbit standard~\cite{IEEE802.3} and BPF (Berkeley Packet Filter) technology~\cite{Li2019}. To exchange the control program with the CCD controller, a subset of protocols of the GigE (gigabit ethernet) Vision 2.0 standard is used~\cite{AIA_GIGE}. Two types of logical channels are implemented---a control channel and a video channel. The control channel is based on the GigE Vision Control Protocol (GVCP). The video channel is a stream of video data received from the system to the computer's internal memory and is based on the GigE Vision Streaming Protocol (GVSP).

The power source generates seven highly stable supply voltages, performs telemetry of these voltages and additionally implements pressure telemetry by connecting a cryostat pressure sensor.

%Winpcap network-based data capture and analysis systems 

%%%%%%%%%%%%%%%%%%%%%%%%%%%%%%%%%%%%%%%%%%
\section{Camera Control Software}

The software for data acquisition, reading and processing~\cite{Afanasieva2015} is programmed using VC++ and QT, and operates under Windows 7/10 x64. The software provides the following capabilities:
\begin{itemize}
\item  Control of the CCD System and exposure parameters' setup;
\item  Visualization and analysis of video data;
\item  Image storage in the Flexible Image Transport System (FITS) standard~\cite{Pence2010};
\item  Interactive and automatic observation modes;
\item  Telemetry and diagnostics of the CCD System;
\item  Software development kit (SDK).
\end{itemize}

A standalone program runs multiple threads in parallel: an interface and service thread, a control thread, a visualization thread and a sorting thread. The main thread provides the user interface and execution of commands. The service thread provides the observation process. The control thread receives packets and retrieves data. The sorting thread forms a frame with an image.

The implemented structure allows you to instantly respond to commands and execute them, while the process of visualization and control is simultaneously carried out independently according to its own timeline.

%%%%%%%%%%%%%%%%%%%%%%%%%%%%%%%%%%%%%%%%%%
\section{Study of the Effect of Charge Spreading}

\begin{figure}[H]
\includegraphics[scale=1.0]{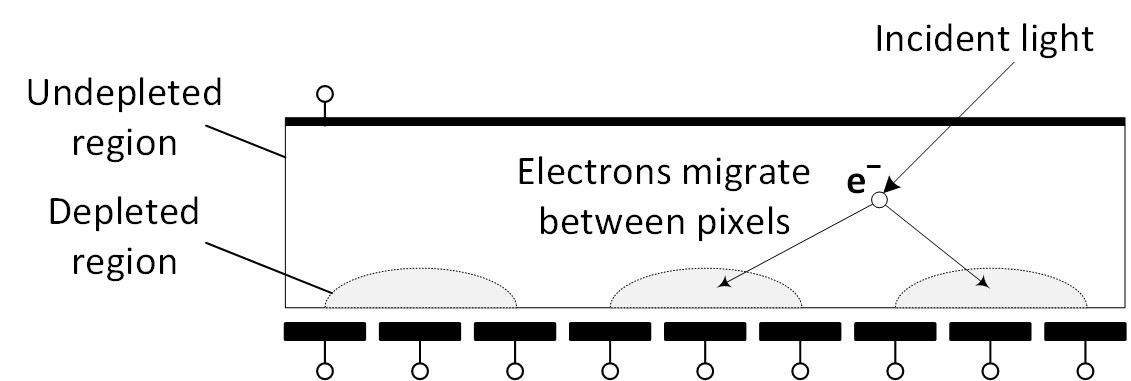}
\caption{Cross-section of the CCD, showing the effect of photoelectron charge collection. \label{fig4}}
\end{figure}

CCD261-84 has a particularly thick ($200\,\upmu$m) high-resistance silicon substrate. Such photodetectors are characterized by the so-called lateral diffusion effect, or charge spreading into adjacent image elements (Figure~\ref{fig4}), leading to deterioration in image quality. To reduce this effect, it is necessary to increase the depth of the substrate depletion region.

The use of High-Rho technology makes it possible to achieve full depletion of a thick high-resistance substrate. This is achieved by applying high negative potential (up to $-$100\,V) to the back side of the device. At the same time, the bias on the front side of the substrate remains the same, maintaining the usual voltage level for the clock phases and the video signal output circuits.

Previously, the effect of charge spreading was investigated in~\cite{Downing2006}. The essence of the estimation method is to calculate the two-dimensional spatial autocorrelation function $R_{m,n}$ of the studied image element $(m,n)$ using the difference $D$ between two uniformly illuminated image arrays of size $M\times N$. The value of the function is calculated by the autocorrelation expression: 
\begin{linenomath}
\begin{equation}
R_{m,n}=
 \frac{\displaystyle 
   \sum_{j=1}^{N-n}\sum_{i=1}^{M-m}D_{i,j}D_{i+m,j+n}}
  {\displaystyle
 \sum_{j=1}^{N}\sum_{i=1}^{M}D^2_{i,j}}.
\end{equation}
\end{linenomath}

\begin{figure}[H]
\includegraphics[width=0.5\textwidth]{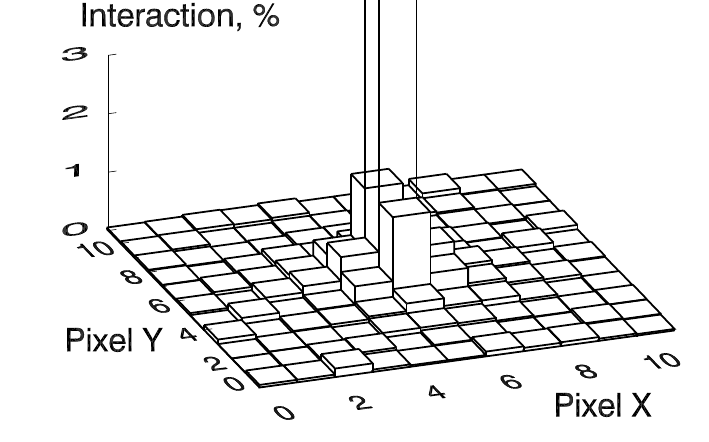}
\includegraphics[width=0.5\textwidth]{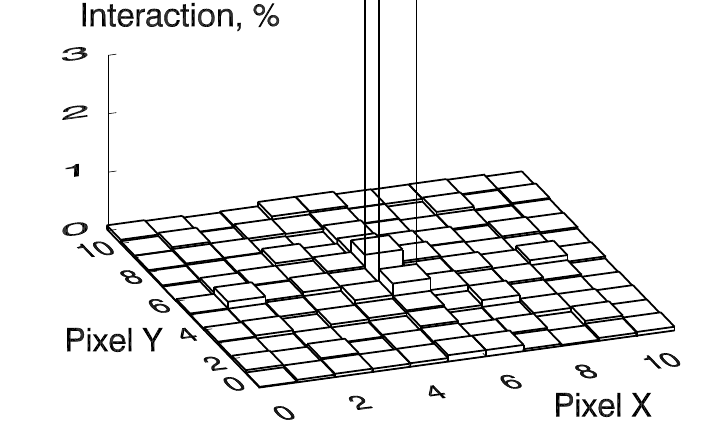}
\caption{Results of the 2D spatial autocorrelation analysis of data taken with CCD261-84 with a median signal value $\thicksim$28,000 ADU:  in low-gain mode (\textbf{left}) and high-gain mode (\textbf{right}). \label{fig5}}
\end{figure}

The left plot of Figure~\ref{fig5} shows the results of the 2D spatial autocorrelation analysis of two flat-field images without binning in low-gain mode with an average flux of $\thicksim$90~ke$^{-}$. %MDPI: please confirm if this is unit? if yes, please change all unit into normal style.
 The central pixel is off-scale and is of value 100\%. The correlation is clearly visible and amounts to 0.64\% in the horizontal and 1.49\% in the vertical direction. The effect of charge spreading in CCD261-84 is lower than in E2V~CCD44-82 with the same flux: 1.4\% in the horizontal and 2.2\%  in the vertical direction (for details, see~\cite{Downing2006}). The right plot of Figure~\ref{fig5} is the result of an analysis of two $\thicksim$17~ke$^{-}$ flat-field images without binning in high-gain mode. The vertical correlation is slightly visible and amounts to 0.35\%.

\begin{figure}[H]
\begin{adjustwidth}{-\extralength}{0cm}
\centering
\includegraphics[width=1.3\textwidth]{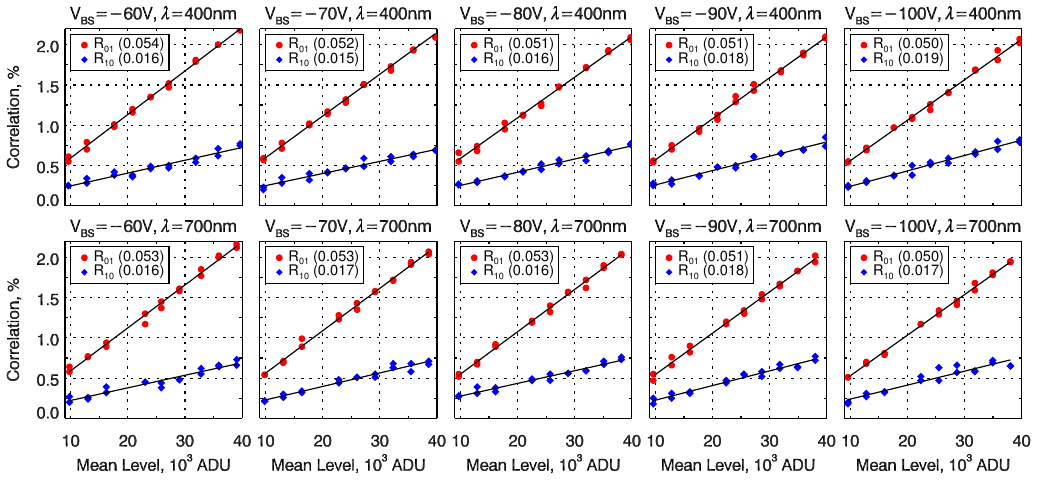}
\end{adjustwidth}
\caption{Measured flat-field correlation coefficients $R_{01}$ and $R_{10}$ for the CCD261-84 in low-gain mode and their best-fit line slope factors (given in parentheses). \label{fig6}}
\end{figure}

By controlling the substrate voltage $V_{bs}$, one can find the optimal value for reducing the lateral charge diffusion. Figure~\ref{fig6} shows the relationship between adjacent image elements, depending on the signal strength at different photodetector back substrate voltages at incident radiation wavelengths of 400 and 700~nm. Testing images were obtained using a flat-field stand, including the CCD System and an integrating sphere calibration standard (Gooch \& Housego). The operating temperature of the photodetector was $-$130 $\pm$  0.1 $^\circ$C.

We see a linear increase in correlation values while increasing the light output. The vertical nearest neighbor coefficient $R_{01}$  is systematically much higher than the other coefficients. This behavior is observed for all values of~$V_{bs}$. The dependence of charge diffusion on voltage is weakly expressed. By lowering the voltage on the substrate to $-$100\,V, we reduce this dependence but within small limits (0.04--0.15\%).

As mentioned above, the CCD controller provides high stability of the video channel transfer characteristic. The transfer characteristic is represented by the gain in the video channel, the so-called charge-to-voltage conversion factor of the CCD system. The conversion factor characterizes the relationship between the charge collected in each pixel and the analog-to-digital unit (ADU) value in the output image~\cite{Howell2006} and is used, in almost all calculations, to obtain real luminous flux values.

Taking into account the summed correlation between pixels, it is possible to more accurately calculate the gain in the system.

%%%%%%%%%%%%%%%%%%%%%%%%%%%%%%%%%%%%%%%%%%
\section{The Index of Dispersion}

Contrary to popular misconceptions, the signal registered by a CCD does not strictly follow Poisson statistics. Regarding astronomical research, the registration of weak signals whose statistics are distorted by the readout noise introduced by electronics is of particular interest.

As a criterion for checking the deviation from the Poisson distribution, the dispersion index, the so-called Fano factor, is effective~\cite{Fano1947}. By definition, the dispersion index is the ratio of the variance of counts to the average value of the registered signal~(concerning the CCD study, the method is described in~\cite{Afanasieva2016}). For Poisson distribution, this ratio is equal to one and corresponds only to a certain range of registered values.

\begin{figure}[H]
\includegraphics[angle=90,width=1.0\textwidth]{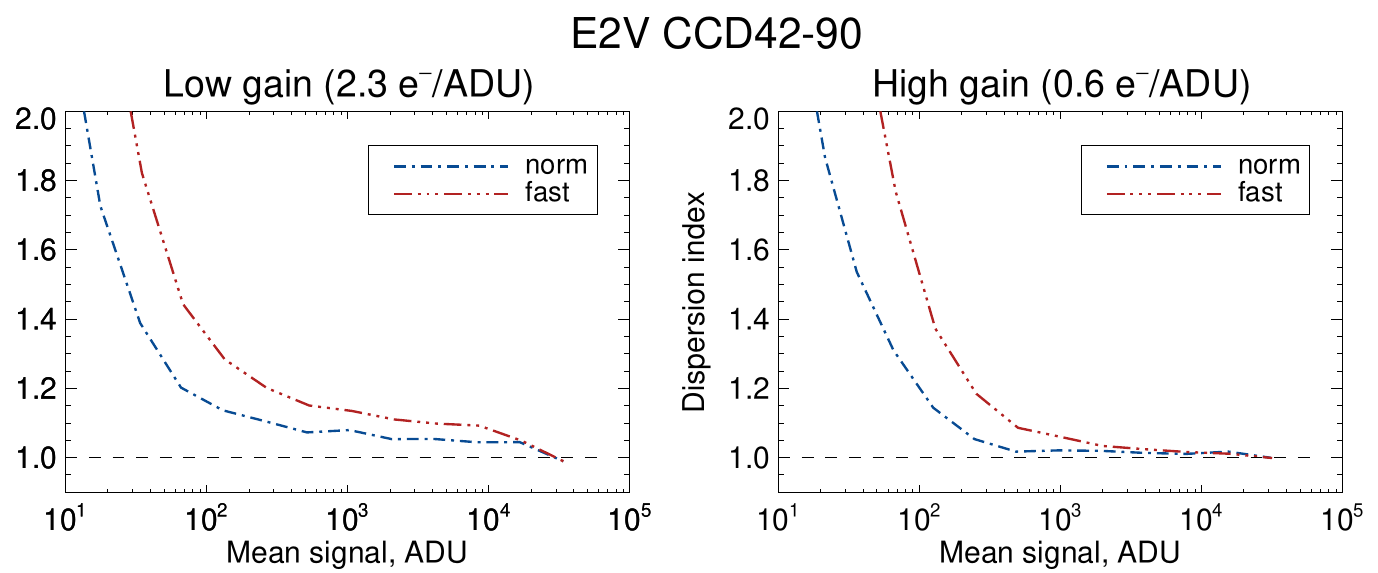}
\includegraphics[angle=90,width=1.0\textwidth]{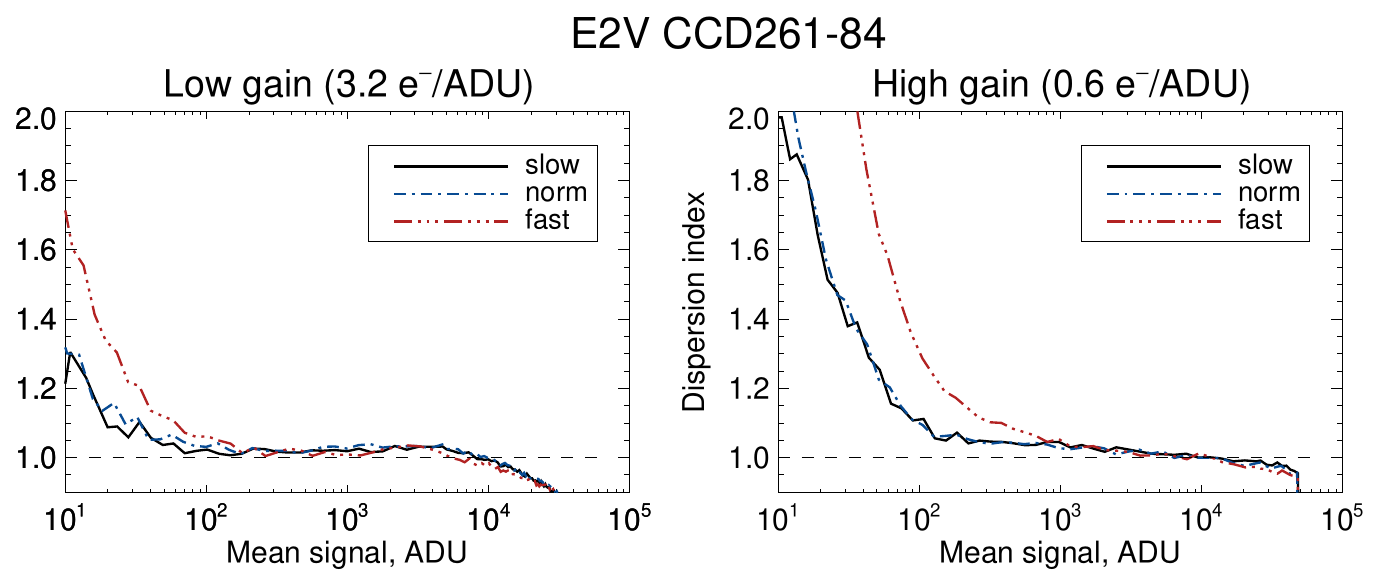}
\caption{Graphs of the dependence of the dispersion index on the flux of the recorded signal in various readout modes for the CCD261-84 (\textbf{bottom}) in comparison with the CCD42-90~(\textbf{top}), according to~\cite{Afanasieva2016}.\label{fig7}}
\end{figure}

Figure~\ref{fig7} shows the dependence of the dispersion index on the magnitude of the registered signal in various modes for CCD261-84 in comparison with E2V~CCD42-90~($2K\times4.5K$ with a pixel size of $13.5\,\upmu$m), the detector previously used with the SCORPIO-2 device. The left and right panels correspond to two gain modes---low ($\times1$) and high ($\times4$), respectively. We see that, compared to the previous detector (CCD42-90), with the new CCD, the Poisson statistics can be accepted for relatively low fluxes---100--200~ADU at the `normal' and `slow' readout rates.

The plots presented allow one to choose the optimal choice of exposure time and readout mode to prevent distortion of count statistics during observations of astrophysical objects with different brightness levels. 

%%%%%%%%%%%%%%%%%%%%%%%%%%%%%%%%%%%%%%%%%%
\section{Defringing}

\begin{figure}[H]
\begin{adjustwidth}{-\extralength}{0cm}
\centering
\includegraphics[width=1.1\textwidth]{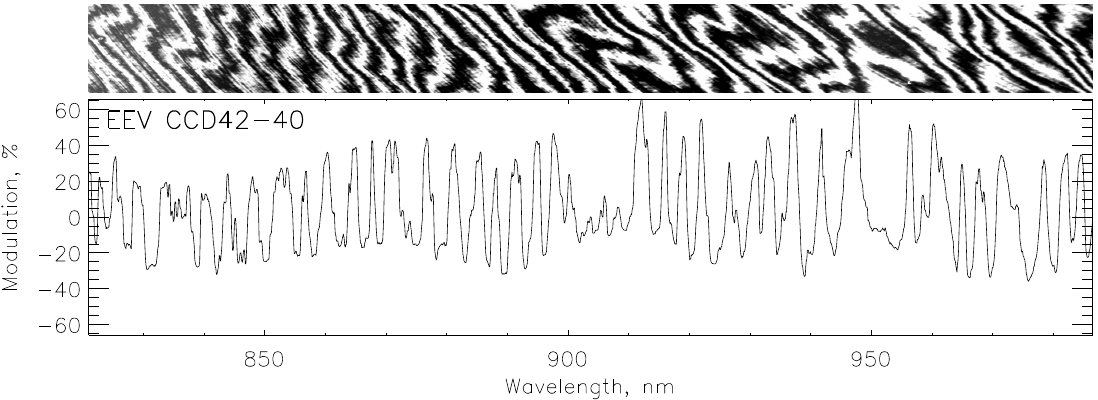}
\includegraphics[width=1.1\textwidth]{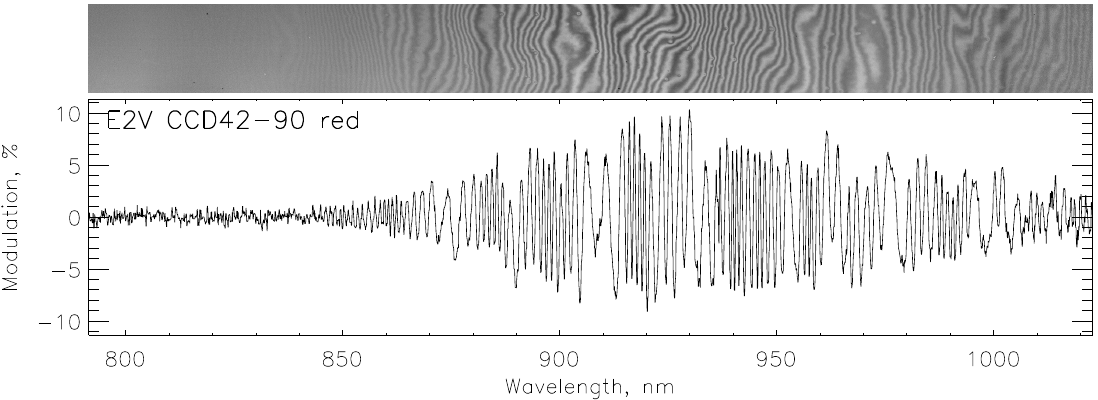}
\includegraphics[width=1.1\textwidth]{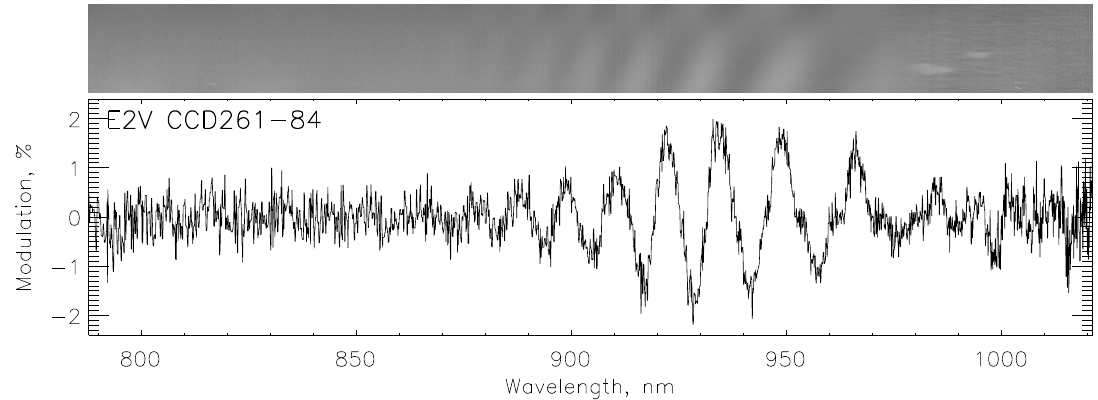}
\end{adjustwidth}
\caption{Pattern of fringes in the red spectral range taken from flat-field observations with CCD42-20 at the SCORPIO~\cite{Afanasiev2005} (\textbf{top}), CCD42-90 (\textbf{middle}), and CCD261-84 (\textbf{bottom}). The last two datasets were obtained with the same grism on {SCORPIO-2} illuminated by a system of LEDs in the standard integration sphere of the standard SCORPIO-2 calibration system~\cite{Afanaisev2017}. The flat field in the archival CCD42-40 was illuminated by a halogen lamp in the previous version of a similar calibration system. Each panel contains a grayscale representation of a fragment of normalized flat field (vertical width is 200 pixels) at a scale of $\pm25$\% of the average intensity. The plot below shows a horizontal cut along this fragment. \label{fig8}}
\end{figure}

One of the disadvantages of BSI CCDs is the effect of the interference of incident and reflected waves in the NIR range, the so-called fringes. The reverse side silicon bulk entering wave may be reflected from the front surface of the substrate and, returning back, interfere with the incident wave. As a result, depending on how many times the wavelength fits on the thickness of the pixel, the interference can either increase or decrease the amplitude of the resulting wave. The dependence of the pixel sensitivity on the wavelength becomes a periodic function (Figure~\ref{fig8}). 

Teledyne E2V uses thicker substrates in its devices to remove this effect. This makes it possible to shift interference effects to wavelengths by about $1\,\upmu$m, where the sensitivity of the silicon substrate is negligible. Additionally, to reduce interference, multilayer coatings of the substrates with various materials are used to prevent wave reflections. Figure~\ref{fig8} clearly demonstrates the very low amplitude of the fringes detected by the new device: it reaches 1--2\% at a wavelength 900--950~nm, while, in detectors of the previous generation, we have significantly higher values (5--10\% for CCD42-90 and even more than 50\% for CCD42-40). The CCD42-90 and CCD42-40 have a basic midband anti-reflection (AR) coating, and the CCD261-84 has a Multi-2 AR coating.

%%%%%%%%%%%%%%%%%%%%%%%%%%%%%%%%%%%%%%%%%%
\section{Observations and Processing}

Since 2020, the camera with CCD261-84 has been used as a detector on the multimode focal reducer {SCORPIO-2}~\cite{SCORPIO2} at the 6 m telescope of the SAO RAS. This system  operates in  different readout modes, depending on the type of observation mode of the focal reducer. The full-format and original binning $1\times1$ is accepted only in the case of spectral observations with the integral-field unit~(SCORPIO-2/IFU, \cite{Afanaisev2018}). In the most commonly used long-slit spectroscopic mode, the full-format CCD is read in $1\times2$ binning mode, which provides an optimal spatial sampling $0\mbox{.\kern -0.7ex\raisebox{.9ex}{\scriptsize$\prime\prime$}}4/$pix along the spectrograph slit. For the same reason, reading with $2\times2$ binning is accepted in the {SCORPIO-2} direct image mode, whereas only the central square fragment ($2048\times2048$ pix in the original binning) is used. In the last case, the detector provides the $6\mbox{.\kern  0.13ex.% 
  \kern -0.95ex\raisebox{.9ex}{\scriptsize$\prime$}% 
  \kern -0.1ex% 
 }8$ field of view with the same spatial scale of $0\mbox{.\kern -0.7ex\raisebox{.9ex}{\scriptsize$\prime\prime$}}4/$pix.  

\begin{figure}[H]
\hspace{-12pt}
\includegraphics[angle=90,width=0.94\textwidth]{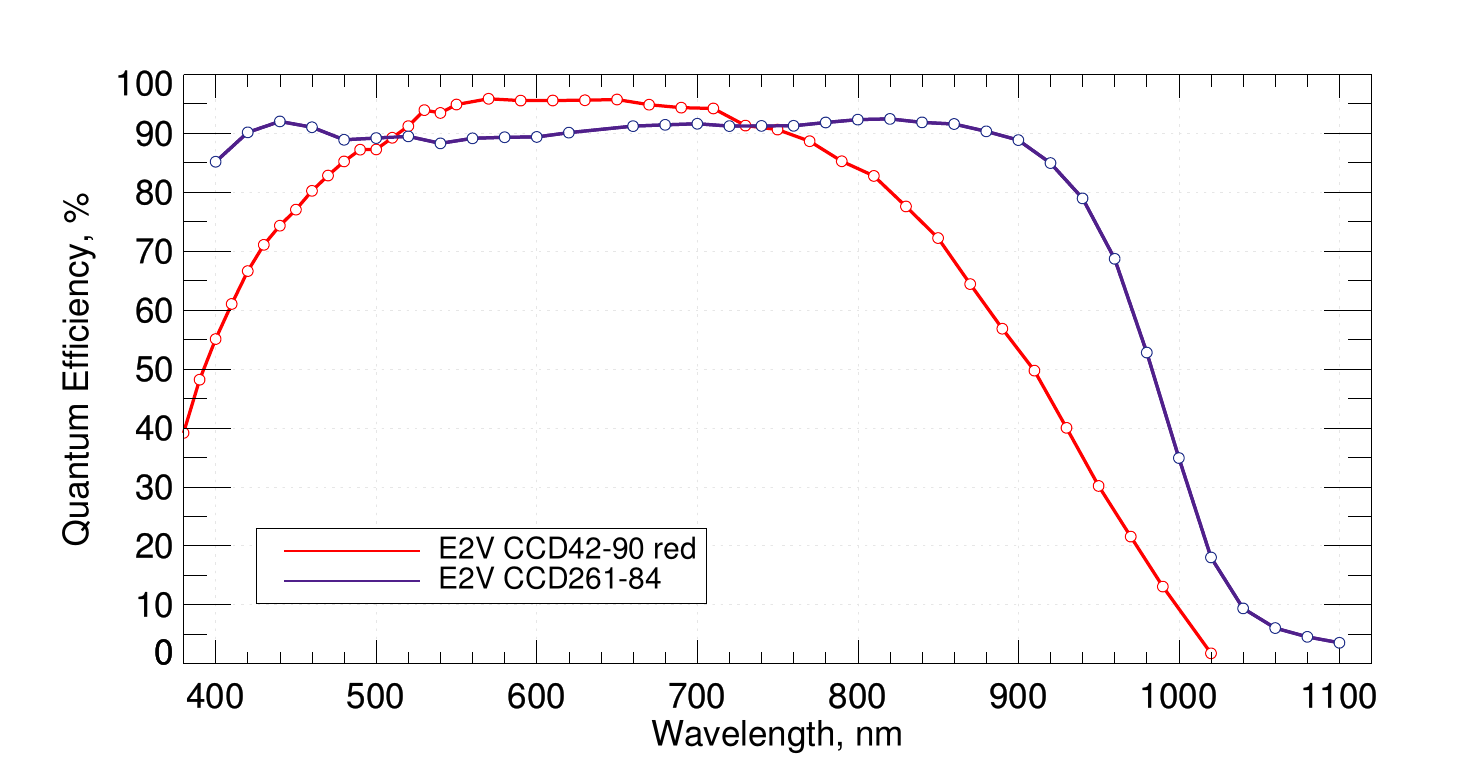}
\caption{Quantum efficiency (QE) of CCD261-84 in comparison with CCD42-90. QE was obtained using an experimental setup which included the explored CCD System, a monochromator, and a rotating mirror. The measurement method is described in~\cite{Jahne2020}. The increased sensitivity in the blue wavelength range of the blue curve is due to Multi-2 AR coating. \label{fig9}}
\end{figure}

Compared to the previous detector CCD42-90, the newest one has significantly higher sensitivity with low contrast of fringes at a wavelength of more than $800$~nm~(Figure~\ref{fig9}). Both of these advantages allow us to use {SCORPIO-2} to solve new observational tasks in the red spectral range, mainly the spectroscopy of objects fainter than the foreground sky emission. Recent examples of this include the measurement of CaII triplet velocities at 845--866~nm in the dwarf galaxy KKH~22 \cite{Karachentsev2020} and spectroscopy of the distant quasar at redshift $z = 5.47$ \cite{Khorunzhev2021}.

\begin{figure}[H]
\begin{adjustwidth}{-\extralength}{0cm}
\centering
\includegraphics[width=1.13\textwidth]{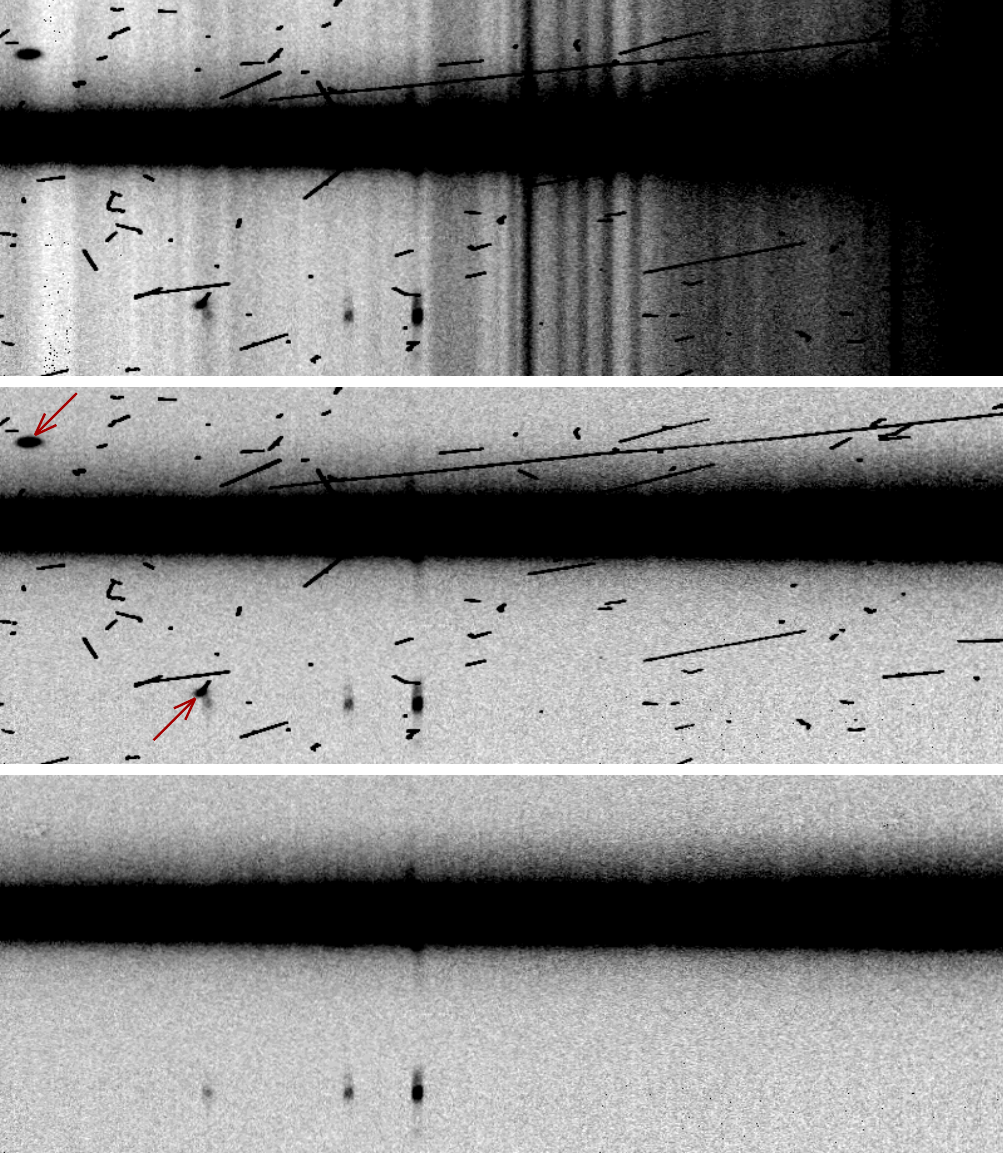}
\end{adjustwidth}
\caption{Long-slit spectrum obtained with {SCORPIO-2} exposed by CCD261-84: $800\times300$ pixel fragment near the [OIII] and  H$\beta$ emission lines of the ionized gas cloud between the galaxies NGC~235 and NGC~232 (for details, see~\cite{Keel2022}). The sum of  $4\times600$~s exposures is shown. From top to bottom: the raw spectrum including airglow emission and cosmic ray hits; sky emission is subtracted, the red arrows show hits after cosmic ray particles are trapped during the readout; the ‘cleaned' spectrum after sigma-filter comparison of individual frames. \label{fig10}}
\end{figure}

The cost of using this high-efficiency CCD is a higher rate of cosmic ray hits (CHs) than with a thin back illuminated detector. Figure~\ref{fig10} shows an example of the CH pattern collected on the small fragment of the CCD261-84 chip during 40 min of exposure: the length of some cosmic ray tracks can reach several hundreds (in some cases, even several thousands) of pixels. Fortunately, most of these high-contrast CHs are easy to remove with  standard algorithms based on Laplacian edge detection (see L.A.COSMIC program in~\cite{lacosmic}). However, some cosmic rays trapped during the readout process introduce a more serious problem because their tracks are smoothed and seem like images of stars or emission knots (the red arrows in Figure~\ref{fig10}, middle). To remove these sorts of CHs in {SCORPIO-2}, a sigma-clipping procedure to multiple (3--5) frames of the same objects is employed in {SCORPIO-2} data reduction after subtraction of the sky emission in individual frames (Figure~\ref{fig10}, bottom). Our first experience showed that new types of algorithms need to be developed in this area, different from the standard ones.

%%%%%%%%%%%%%%%%%%%%%%%%%%%%%%%%%%%%%%%%%%
\section{Results}

We have presented the characteristics and performance of the CCD261-84 camera system installed on the multimode focal reducer {SCORPIO-2} at the 6 m telescope of the SAO RAS. The main performance characteristics of the CCD system with the CCD261-84 are shown in Table~\ref{tab1}.

\begin{table}[H] 
\caption{Characteristics of the CCD261-84 camera system.\label{tab1}}
\newcolumntype{C}{>{\centering\arraybackslash}X}
\begin{tabularx}{\textwidth}{CCC}
\toprule
						&  \multicolumn{2}{c}{ \textbf{Value} } 		\\
		\noalign{\smallskip} \cline{2-3} \noalign{\smallskip}
\textbf{Parameter	}				&\textbf{ Low Gain  	}		& \textbf{High Gain} 	\\
\midrule
CCD coating type				&  \multicolumn{2}{c}{Astro Multi-2}  	\\
Image size (mm)				&  \multicolumn{2}{c}{ $30.7\times61.6$ }  \\
Pixel pitch ($\upmu$m) 			&  \multicolumn{2}{c}{ $15\times15$ } 	\\
Readout rates (kpixel/s)		& \multicolumn{2}{c}{65 (slow), 185 (normal), 335 (fast)}  \\
Readout noise ($e^{-}$)		& 2.62~@ 65~kpixel/s	&  2.18~@~65~kpixel/s \\
Gain factor (e$^{-}$/ADU)		& 3.2  	& 0.6  			\\
Full well capacity (ke$^{-}$/pixel)		& 165  	& 35 				\\
Dark current (e$^{-}$/s/pixel)		& \multicolumn{2}{c}{ 0.001 }		\\
Operating temperature ($^\circ$C) & \multicolumn{2}{c}{ $-130$ } 	\\
Thermal stability ($^\circ$C) 	& \multicolumn{2}{c}{ $\pm 0.1$ } 	\\
Linearity (\%)   				&  \multicolumn{2}{c}{$>$ 99.3~(0--140~ke$^{-}$) }	\\
\bottomrule
\end{tabularx}
\end{table}

%The quality of spectral images has been improved by reducing the lateral charge diffusion.
%This CCD system is more powerful than the CCD42-90 camera system that has been in use earlier, with QE of >90\% from 450--900 nm and , factors of a few to tens improvement at  long wavelengths.

The operating spectral range of the SCORPIO-2 multimode focal reducer, which provides a significant part of the photometric and spectral observational programs of the BTA, has been extended to the red region (up to 1000~nm).

For the designed CCD system, a deviation from the Poisson statistics is observed for fluxes weaker than 250~e$^{-}$ with a readout noise of 2.18~e$^{-}$. The CCD system can be regarded as almost `ideal' in the flux range of 250--40,000~e$^{-}$. The dark current decreased by a factor of three compared to the previous detector (CCD42-90). The amplitude of the fringes decreased five times.

An image filtering procedure has been developed to remove traces of cosmic particles in relation to detectors with a thick substrate.

The features and size of fringes in a thick substrate have been studied, and recommendations have been developed for taking them into account in observations.

%%%%%%%%%%%%%%%%%%%%%%%%%%%%%%%%%%%%%%%%%%
\section{Discussion}

This article presents a finished working CCD system based on a fully depleted scientific detector CCD261-84 with substrate thickness of $200\,\upmu$m. Descriptions of implemented CCD systems with such a photodetector have not been published before; there are only data on the study of the photodetectors themselves by the manufacturer~\cite{Robbins2011, Jorden2010, Jorden2014} and measurements of the detector characteristics to substantiate the charge transfer model in such detectors, as well as to determine the expediency of using detectors in planned projects~\cite{Weatherill2016}. It is shown that the thoughtful design of the camera and the use of a universal CCD controller developed at the SAO RAS made it possible to achieve high photometric characteristics that are inherent in the CCD detector itself. Primarily, it has low readout noise and low dark current, which makes it possible to effectively use the device for spectral observations of faint objects with exposures up to 1 h. The achieved readout noise was 2.18~e$^{-}$~@~65~kpixel/s at the declared value of 2.8~e$^{-}$~@~50~kpixel/s in the manufacturer datasheet. Such a level of readout noise is inherent in a few working astronomical CCD systems.

To characterize the degree of charge spreading in the substrate, we used the method for finding the two-dimensional autocorrelation function of the difference between two flat-field frames. The application of this method does not require the construction of a complex stand with a movable point light source followed by obtaining a point spread function (PSF); rather, only a flat-field stand is needed. In this case, the method makes it possible to estimate the degree of charge spreading for a given wavelength and a known charge value in a pixel. This article shows that in the operating range of substrate voltage ($-$60\,V{\ldots}$-$100\,V), the amount of charge spreading varies slightly depending on the substrate voltage. This is consistent with the data of Figure~4 in~\cite{Robbins2011}, which shows that the PSF changes insignificantly in this range of substrate voltages for samples of devices with varying degrees of silicon depletion. When choosing the operating voltage of the back substrate, it is also necessary to consider the fact that, with an increase in the potential difference, the generation of a parasitic charge in hot pixels increases (Figure 8 in~\cite{Robbins2011}). We also demonstrated a pronounced direct dependence of the autocorrelation coefficients on the signal level, which is consistent with theory. We can compare our data with the data from Figure 6.9 in~\cite{Weatherill2016}. In the range of signals from 20 to 1000~ke$^{-}$ at a substrate voltage of $-$70\,V,  the author does not observe a clear dependence, although the order of the coefficients is the same. This may indicate that the quality of the video channel in the data acquisition system used.

The Fano factor also indicates the quality of the video signal processing channel. The graph shown for this system (Figure~\ref{fig7}, bottom) testifies to the high quality of the video channel, evidenced by the absence of interference and distortion of statistics in various input signals. Since this parameter is practically not given in the description of CCD systems, it is difficult to compare it with other projects.

The characteristic quantum sensitivity curve for detectors with depleted silicon and the Multi-2 anti-reflection coating is also given in Figure 14 in~\cite{Robbins2011}. The measured values of the quantum efficiency for our CCD system are in good agreement with the curve from this publication as well as with the typical curve presented in the datasheet of the detector, considering some losses in the camera optical window. Thanks to the Multi-2 coating, the detector has very high sensitivity, not only in the red, but also in the blue region of the~spectrum.

%%%%%%%%%%%%%%%%%%%%%%%%%%%%%%%%%%%%%%%%%%
\section{Conclusions}

In this paper, we describe the design, implementation and operating principles of an astronomical camera system based on a large-format CCD261-84 detector. To use the new BSI CCD with a very thick substrate, a CCD controller with the possibility of generating a controlled high-voltage level has been developed at SAO RAS. The CCD System with CCD261-84 is now used as a detector on the multimode focal reducer {SCORPIO-2} at the 6 m telescope of the SAO RAS. Studies of the photometric characteristics of the system as well as the effect of charge spreading in the substrate and the formation of fringes have been carried out. Low readout noise and dark current were achieved. Based on the images obtained in the observations, new methods of processing and reduction of spectral images have been developed to minimize the increased impact of cosmic ray particles. The developed CCD System has significantly higher sensitivity, with a low contrast of fringes at a wavelength of more than 800~nm, which makes it possible to use the {SCORPIO-2} at the SAO RAS 6 m telescope to solve new observational tasks in the red spectral range.

%%%%%%%%%%%%%%%%%%%%%%%%%%%%%%%%%%%%%%%%%%
\vspace{6pt} 

%%%%%%%%%%%%%%%%%%%%%%%%%%%%%%%%%%%%%%%%%%
%% optional
%\supplementary{The following supporting information can be downloaded at:  \linksupplementary{s1}, Figure S1: title; Table S1: title; Video S1: title.}


%%%%%%%%%%%%%%%%%%%%%%%%%%%%%%%%%%%%%%%%%%
\authorcontributions{Conceptualization, V.M., V.A., I.A. and N.I.; methodology, all authors; software, I.A., V.M. and A.M.; validation, V.M., I.A., V.A., N.I. and M.P.; data curation, V.M., I.A., A.M. and E.M.; writing---original draft, V.A.; writing---review \& editing, V.A., I.A., V.M., A.M., E.S. and E.M.; visualization, I.A., V.A., A.M. and E.M.; supervision, I.A. and V.M.; project administration, I.A. All authors have read and agreed to the published version of the manuscript.}

\funding{The observational data were collected using the unique scientific facility Big Telescope Alt-azimuthal of SAO RAS, and data processing was performed with the financial support of grant No075-15-2022-262 (13.MNPMU.21.0003) of the Ministry of Science and Higher Education of the Russian Federation.}

\institutionalreview{Not applicable.}

\informedconsent{Not applicable.}

\dataavailability{The data are available from the first author upon reasonable request.} 

\acknowledgments{We express our gratitude to the staff of the Laboratory of Spectroscopy and Photometry of Extragalactic Objects (SAO RAS): Perepelitsyn,~A.E.; Uklein,~R.I.; Oparin,~D.V.; and Kotov,~S.S. for their assistance in laboratory measurements of CCD characteristics.}

\conflictsofinterest{The authors declare no conflicts of interest.} 

%%%%%%%%%%%%%%%%%%%%%%%%%%%%%%%%%%%%%%%%%%
%% Optional
%\sampleavailability{Samples of the compounds ... are available from the authors.}

%% Only for journal Encyclopedia
%\entrylink{The Link to this entry published on the encyclopedia platform.}

%%%%%%%%%%%%%%%%%%%%%%%%%%%%%%%%%%%%%%%%%%
\begin{adjustwidth}{-\extralength}{0cm}

\reftitle{References}

\end{adjustwidth}
\end{document}